# A new fMRI data analysis method using cross validation: Negative BOLD responses may be the deactivations of interneurons.

Hiroshi Tsukimoto and Takefumi Matsubara

Tokyo Denki University

tsukimoto@c.dendai.ac.jp

December 10, 2017

**Abstract**

Although functional magnetic resonance imaging (fMRI) is widely used for the study of brain functions, the blood oxygenation level dependent (BOLD) effect is incompletely understood. Particularly, negative BOLD responses(NBRs) is controversial. This paper presents a new fMRI data analysis method, which is more accurate than the typical conventional method. The authors conducted the experiments of simple repetition, and analyzed the data by the new method. The results strongly suggest that the deactivations(NBRs) detected by the new method are the deactivations of interneurons, because the deactivation ratios obtained by the new method approximately equals the deactivation ratios of interneurons obtained by the study of interneurons. The (de)activations detected by the new method are largely different from those detected by the conventional method. The new method is more accurate than the conventional method, and therefore the (de)activations detected by the new method may be correct and the (de)activations detected by the conventional method may be incorrect. A large portion of the deactivations of inhibitory interneurons is also considered to be activations. Therefore, the right-tailed t-test, which is usually performed in the conventional method, does not detect the whole activation, because the right-tailed t-test only detects the activations of excitatory neurons, and neglect the deactivations of inhibitory interneurons. A lot of fMRI studies so far by the conventional method should be re-examined by the new method, and many results obtained so far will be modified.

## 1. Introduction

Functional magnetic resonance imaging (fMRI) is widely used for the study of human brain functions. fMRI is based on the blood oxygenation level dependent (BOLD) effect(1). The BOLD response is divided into positive BOLD response (PBR) and negative BOLD response (NBR). In particular, NBR remains incompletely understood(2,3,4).

The cause of the problem of NBR may be an inappropriate analysis method. In the typical conventional fMRI data analysis method, the regression analysis of each voxel is performed. T-tests are performed on the coefficients of regression formulas. Multiple comparison adjustments are performed in many cases(5). In the multiple comparison adjustment, the threshold for p value is modified. The regression analysis of several adjacent voxels is expected to work better than the regression analysis of one voxel, because voxels are connected with each other in reality.

This paper presents a new fMRI data analysis method. In the conventional method, tasks/rests are independent variables and fMRI values are dependent variables. In the new method, contrary to the conventional method, tasks/rests are dependent variables and fMRI values are independent variables. By this exchange, multiple regression analysis with several voxels, which properly models the situation that voxels are connected with each other, is possible. The regression analysis with several voxels is expected to be more accurate than the conventional method and is also expected to solve the problem of NBRs.

In the new method, cross validations are performed to estimate the accuracies of regression formulas with several variables. Regression formulas with small mean squared prediction errors (MSPEs) are selected in an appropriate method, and t-tests (two-tailed) are performed on the coefficients of the regression formulas. Positive coefficients that are statistically significant show PBRs (activations), and negative coefficients that are statistically significant show NBRs (deactivations).

The authors conducted the experiments of simple repetition using fMRI. It is well known that the temporal lobe and the motor cortex are activated in simple repetition(6). The new method was applied to the temporal lobe and the motor cortex. The new method with 8 voxels is more accurate than the conventional method. The deactivation ratio $(= \frac{\text{the number of deactivated voxels}}{\text{the number of deactivated voxels} + \text{the number of activated voxels}})$ of the temporal lobe is approximately 30%-33%. Neurons in the brain are broadly divided into excitatory neurons (e.g. pyramidal neurons) which release glutamate, and interneurons which are primarily inhibitory and release GABA. It was reported that the ratio of the population of interneurons in the human temporal lobe is approximately 37.7%(7). We have to consider the energy consumption of interneurons and the ratio of deactivated interneurons in activation areas, the both of which are not well known. However, under an appropriate assumption, the ratio of deactivations of interneurons is approximately 25.4%-33.8%.   As the ratio of deactivations obtained from the experiments is 30%-33%, the deactivations obtained from the experiments are probably the deactivations of interneurons.

As described above, the new method found that NBRs are the deactivations of interneurons, which could not be found by the conventional method, because the conventional method is inaccurate. Metaphorically speaking, the conventional method is a microscope with magnification of 10 times and the new method is a microscope with magnification of 20 times. The deactivations of interneurons, which cannot be detected by a microscope with 10 times (the conventional method), can be detected by a microscope with magnification of 20 times (the new method).

The (de)activations detected by the conventional method is largely different from those detected by the new method. The conventional method is less accurate than the new method, and therefore the (de)activations detected by the conventional method may be incorrect and the (de)activations detected by the new method may be correct. A large portion of the deactivations of inhibitory interneurons is also considered to be activations. Therefore, the right-tailed t-test, which is usually performed in the conventional method, does not detect the whole activation, because the right-tailed t-test only detects the activations of excitatory neurons, and neglect the deactivations of inhibitory interneurons. A lot of fMRI studies thus far by the conventional method should be re-examined by the new method, and many results obtained thus far will be modified.

## 2. A new fMRI data analysis method
## 2.1 The features of the new fMRI data analysis method
1. fMRI values are independent variables and tasks/rests are dependent variables.

In the conventional method, tasks/rests(convolved with hemodynamic response function:HRF) are independent variables and fMRI values are dependent variables. In the new method, fMRI values are independent variables and tasks/rests (convolved with HRF) are dependent variables. By this exchange, multiple regression analysis with several voxels is possible.

2. Multiple regression analysis with several voxels

First, we explain the case of one voxel. The regression formula is as follows:

$$y = ax + b,$$

where y stands for task(1)/rest(0) (As convolved with HRF, the range of $y$ is a little wider than the 0-1 range), $x$ stands for a fMRI value, $a$ stands for the coefficient and $b$ stands for the bias. The above formula is a prediction model, where a fMRI value predicts task/rest. This model does not show the physical causal relationship. This model shows the correlation between fMRI values and tasks/rests. It can be expected that the regression analysis with several voxels is more accurate than the regression analysis with one voxel, because voxels are connected with each other and voxels do not work independently. The typical simple set of several voxels is a cube with a side of 2 voxels. The total number of the voxels is 8(=2×2×2). The regression formula with 8 voxels is as follows:

$$y = a_1 x_1 + \cdots + a_8 x_8 + b,$$

where $y$ stands for task(1)/rest(0) and $x_i$s stand for fMRI values. The above regression analyses are performed for all the cubes. In order to simplify the explanation, we explain the case of two dimensions, where squares with a side of two voxels are considered. Fig.1 shows 16 voxels. The first square is (1,2,5,6), the second square is (2,3,6,7), the third square is (3,4,7,8), the fourth square is (5,6,9,10), and so on. Two examples of the regression formulas are as follows:

$$y = a_1 x_1 + a_2 x_2 + a_3 x_5 + a_4 x_6 + b_1,$$
$$y = a_5 x_2 + a_6 x_3 + a_7 x_6 + a_8 x_7 + b_2.$$

The first formula corresponds to the first square (1,2,5,6), and the second formula corresponds to the second square (2,3,6,7).

| 1 | 2 | 3 | 4 |
|---|---|---|---|
| 5 | 6 | 7 | 8 |
| 9 | 10 | 11 | 12 |
| 13 | 14 | 15 | 16 |

**Figure 1　16 voxels in the case of two dimensions**

## 2.2 The procedures of the new method

The procedures of the new method is explained with the two-dimensional example of Fig.1, for simplification. Consider the regression analysis of four(=2×2) voxels. There are nine

squares.((1,2,5,6),(2,3,6,7),(3,4,7,8),(5,6,9,10),(6,7,10,11),(7,8,11,12),(9,10,13,14),(10,11,14,15), (11,12,15,16))

1. Perform the cross-validations.

Let us assume that the regression formulas and the MSPEs are obtained as shown in Table 1.

**Table 1　　Regression formulas**

| No. | Regression formulas | MSPEs |
| --- | --- | --- |
| 1 | $y = -0.01x_1 + 0.61x_2 + 0.82x_5 + 0.27x_6 + 0.41$ | 0.11 |
| 2 | $y = 0.76x_2 + 0.06x_3 + 0.91x_6 - 0.02x_7 + 0.52$ | 0.08 |
| 3 | $y = -0.01x_3 + 0.06x_4 - 0.07x_7 - 0.03x_8 + 0.45$ | 0.20 |
| 4 | $y = 0.94x_5 + 0.64x_6 - 0.08x_9 - 0.72x_{10} + 0.48$ | 0.09 |
| 5 | $y = 0.70x_6 - 0.06x_7 - 0.88x_{10} - 0.02x_{11} + 0.46$ | 0.15 |
| 6 | $y = 0.01x_7 - 0.06x_8 + 0.08x_{11} - 0.05x_{12} + 0.50$ | 0.18 |
| 7 | $y = 0.04x_9 - 0.86x_{10} + 0.09x_{13} - 0.03x_{14} + 0.48$ | 0.19 |
| 8 | $y = -0.95x_{10} + 0.04x_{11} + 0.08x_{14} - 0.01x_{15} + 0.49$ | 0.16 |
| 9 | $y = 0.01x_{11} + 0.06x_{12} - 0.08x_{15} - 0.02x_{16} + 0.51$ | 0.17 |

2. Sort the regression formulas in ascending order of their MSPEs.

The regression formulas are sorted in ascending order of their MSPEs. See Table 2.

**Table 2　Sorted formulas**

| No. | Regression formulas | MSPEs |
| --- | --- | --- |
| 2 | $y = 0.76x_2 + 0.06x_3 + 0.91x_6 - 0.02x_7 + 0.52$ | 0.08 |
| 4 | $y = 0.94x_5 + 0.64x_6 - 0.08x_9 - 0.72x_{10} + 0.48$ | 0.09 |
| 1 | $y = -0.01x_1 + 0.61x_2 + 0.82x_5 + 0.27x_6 + 0.41$ | 0.11 |
| 5 | $y = 0.70x_6 - 0.06x_7 - 0.88x_{10} - 0.02x_{11} + 0.46$ | 0.15 |
| 3 | $y = -0.01x_3 + 0.06x_4 - 0.07x_7 - 0.03x_8 + 0.45$ | 0.16 |
| 6 | $y = 0.01x_7 - 0.06x_8 + 0.08x_{11} - 0.05x_{12} + 0.50$ | 0.17 |
| 7 | $y = 0.04x_9 - 0.86x_{10} + 0.09x_{13} - 0.03x_{14} + 0.48$ | 0.18 |
| 8 | $y = -0.95x_{10} + 0.04x_{11} + 0.08x_{14} - 0.01x_{15} + 0.49$ | 0.19 |
| 9 | $y = 0.01x_{11} + 0.06x_{12} - 0.08x_{15} - 0.02x_{16} + 0.51$ | 0.20 |

3. Perform t-tests on the coefficients of the regression formulas in ascending order of their MSPEs.

T-tests are performed on the coefficients of the regression formulas in ascending order of their MSPEs. Positive coefficients that are statistically significant show PBRs(activations), and negative coefficients that are statistically significant show NBRs(deactivations).

Table 3  T-tests

| No. | Regression formulas | Significant Voxels I | Significant Voxels II | MSPEs |
|---|---|---|---|---|
| 2 | $y = 0.76x_2 + 0.06x_3 + 0.91x_6 - 0.02x_7 + 0.52$ | $x_2, x_6$ | $x_2, x_6$ | 0.08 |
| 4 | $y = -0.94x_5 + 0.64x_6 - 0.08x_9 - 0.11x_{10} + 0.48$ | $x_5, x_6$ | $x_5$ | 0.09 |
| 1 | $y = -0.01x_1 + 0.61x_2 - 0.82x_5 + 0.13x_6 + 0.41$ | $x_2, x_5$ | | 0.11 |
| 5 | $y = 0.70x_6 - 0.06x_7 - 0.88x_{10} - 0.02x_{11} + 0.46$ | $x_6, x_{10}$ | $x_{10}$ | 0.16 |
| 3 | $y = -0.01x_3 + 0.06x_4 - 0.07x_7 - 0.03x_8 + 0.45$ | | | 0.17 |
| 6 | $y = 0.01x_7 - 0.06x_8 + 0.08x_{11} - 0.05x_{12} + 0.50$ | | | 0.18 |
| 7 | $y = 0.04x_9 - 0.86x_{10} + 0.09x_{13} - 0.03x_{14} + 0.48$ | $x_{10}$ | | 0.19 |
| 8 | $y = -0.95x_{10} + 0.04x_{11} + 0.07x_{14} - 0.01x_{15} + 0.49$ | $x_{10}$ | | 0.20 |
| 9 | $y = 0.01x_{11} + 0.06x_{12} - 0.08x_{15} - 0.91x_{16} + 0.51$ | $x_{16}$ | $x_{16}$ | 0.25 |

Let us assume that significant voxels are obtained as shown in "Significant Voxels I" in Table3.

For example, $x_6$(No.6 voxel) is statistically significant in No.2 formula, No.4 formula and No.5 formula. In such a case, $x_6$ in No.2 formula is selected, and $x_6$ in No.4 formula and No.5 formula are neglected. The same applies to other voxels(variables). Generally, a voxel that turned out to be significant by t-test in a regression formula, is neglected afterwards. The result of this processing is shown in "Significant Voxels II". "Significant Voxels I" stands for the results that do not consider double check. "Significant Voxels II" stands for the results that consider double check.

4. Select reliable significant voxels

Regression formulas whose MSPEs are large, are unreliable, and the significant voxels in regression formulas whose MSPEs are large, are also unreliable. For example, in No.9 regression formula, $x_{16}$ is significant, which is unreliable, because the MSPE of No.9 regression formula is large (0.25). See Table 3. Significant voxels in the regression formulas whose MSPEs are large (= unreliable significant voxels) should be excluded, and only significant voxels in the regression formulas whose MSPEs are small (=reliable significant voxels) should be selected. There are a few method to select reliable significant voxels.

One method is to select significant voxels in the regression formulas whose MSPEs are smaller than a threshold. For example, let the threshold be 0.15 in Table 3, then $x_2$, $x_6$ and $x_5$ are selected. Another method is to select significant voxels up to a certain number. For example, let a certain number be 4 in Table 3, then $x_2$, $x_6$, $x_5$ and $x_{10}$ are selected. As far as we have experienced, the latter method worked better than the former method, and therefore we adopt the latter method, in this paper.

The actual algorithm is as follows:
①Perform t-tests on the coefficients of the regression formulas in ascending order of their MSPEs .
②Do not perform t-tests on the coefficients of the voxels that turned out to be significant before that.
③ Perform t-tests until the number of significant voxels reaches a certain number.

In the above argument, a two dimensional example was explained for simplification. In reality, the regression analysis with 8(=2×2×2) voxels are performed. The regression analysis with 27(=3×3×3) voxels are also performed later.

## 3. Results
### 3.1 The new method is accurate.

As the ratio of the population of interneurons in the human temporal lobe was reported(7), we conducted the block design experiments of simple repetition that activates the temporal lobe including the auditory cortex(6). In the tasks, the participants overtly spoke the sentences that they heard through a headphone. The temporal lobe (Brodmann Area 20,21,22,37,38,41, and 42) was analyzed by the conventional method, and the new method(1 voxel, 8 voxels, and 27 voxels). The regression formulas were sorted in ascending order of their MSPEs. Table 4 shows MSPEs. Values in the table are shown with three significant digits. The same applies to other tables.

Table 4    MSPEs of four methods

| No. | 200th MSPEs | | | | 300th MSPEs | | | |
|---|---|---|---|---|---|---|---|---|
| | 1V(cnv) | 1V | 8V | 27V | 1V(cnv) | 1V | 8V | 27V |
| 1 | 0.564 | 0.0974 | 0.0621 | 0.0652 | 0.626 | 0.111 | 0.0707 | 0.0729 |
| 2 | 0.669 | 0.107 | 0.0810 | 0.0817 | 0.727 | 0.113 | 0.0835 | 0.0855 |
| 3 | 0.790 | 0.0935 | 0.0646 | 0.0690 | 0.888 | 0.108 | 0.0684 | 0.0732 |
| 4 | 1.78 | 0.0958 | 0.0672 | 0.0684 | 1.91 | 0.105 | 0.0724 | 0.0724 |
| 5 | 0.635 | 0.132 | 0.0871 | 0.0844 | 0.678 | 0.143 | 0.0926 | 0.0908 |
| 6 | 1.03 | 0.0723 | 0.0547 | 0.0569 | 1.11 | 0.0801 | 0.0578 | 0.0595 |
| 7 | 0.571 | 0.116 | 0.0807 | 0.0769 | 0.615 | 0.130 | 0.0858 | 0.0846 |
| 8 | 0.450 | 0.122 | 0.0785 | 0.0789 | 0.477 | 0.128 | 0.0884 | 0.0857 |
| 9 | 0.988 | 0.154 | 0.110 | 0.114 | 1.10 | 0.165 | 0.118 | 0.124 |
| 10 | 1.01 | 0.158 | 0.108 | 0.111 | 1.10 | 0.170 | 0.117 | 0.118 |
| Avg. | 0.849 | 0.115 | 0.0794 | 0.0806 | 0.923 | 0.125 | 0.0855 | 0.0867 |

The left half of the table shows the MSPEs of the 200th regression formulas, and the right half of the table shows the MSPEs of the 300th regression formulas. "No." stands for the participant number. "1V (cnv)" stands for the conventional method. "1V", "8V" and "27V" stand for the new method with 1 voxels, 8 voxels and 27 voxels, respectively. The new method(8V or 27V) is more accurate than the new method(1V) and the conventional method.

The MSPEs of the conventional method are large compared with those of the new method(1V, 8V or 27V), because the dependent variables in the conventional method are fMRI values and the dependent variables in the new method are tasks/rests convolved with HRF. Therefore, the comparison between the MSPEs of the conventional method and the MSPEs of the new method makes no sense. We want to check if the new method(8V or 27V) is more accurate than the conventional method. However, as explained above, the comparison between the conventional method and the new method(8V or 27V) makes no sense.

The regression ability of the conventional method is almost the same as that of the new method(1V), because the latter is obtained from the former by exchanging the independent variable and the dependent variable. More detailed explanation is as follows. The coefficient of determination is used as a measure of the regression ability of a regression formula(8). There are several definitions of the coefficient of determination, one of which is equal to the squared correlation coefficient of the dependent variable and the independent variable. As the new method(1V) is obtained from the conventional method by exchanging the independent variable and the dependent variable, the squared correlation coefficient of the dependent variable and the independent variable (=a coefficient of determination) of the conventional method is equal to that of the new method(1V). Therefore, the regression ability of the conventional method is almost the same as that of the new method(1V). Let us check if the new method(8V or 27V) is more accurate than the new method(1V) instead of checking if the new method(8V or 27V) is more accurate than the conventional method.

Table 4 shows that the new method(8V) is more accurate than the new method(1V), and the accuracy of the new method(8V) is almost the same as that of the new method(27V). The above assertion is confirmed quantitatively by the analyses of variance (ANOVA) of MSPEs. Table 5 shows ANOVA of MSPEs in Table4.

**Table 5    The ANOVAs of MSPEs in Table 4**

| 200/300 | A | B | VR (TL) | VR(MC) | FCV | DF1 | DF2 |
|---|---|---|---|---|---|---|---|
| 200 | 1V | 8V | 11.5 | 27.4 | 4.41 | 1 | 18 |
| 200 | 8V | 27V | 0.0225 | 0.185 | 4.41 | 1 | 18 |
| 200 | 1V | 27V | 10.6 | 35.0 | 4.41 | 1 | 18 |
| 300 | 1V | 8V | 13.5 | 63.0 | 4.41 | 1 | 18 |
| 300 | 8V | 27V | 0.0178 | 0.183 | 4.41 | 1 | 18 |
| 300 | 1V | 27V | 12.5 | 45.8 | 4.41 | 1 | 18 |

"200/300" stands for the 200th regression formula or the 300th regression formula.  "A" and "B" stand for the two groups of ANOVA. VR, TL, MC, FCV, DF1 and DF2 stand for variance ratio (=F value), temporal lobe, motor cortex, F-critical value at 0.05, degree of freedom between groups, and degree of freedom within groups, respectively. The variance ratios between the new method(1V) and the new method(8V), and those between the new method(1V) and the new method(27V) are greater than F-critical value. The variance ratios between the new method(8V) and the new method(27V) are less than F-critical value.

We also analyzed the motor cortex (Brodmann Area 4 and 6), and obtained the similar results to the temporal lobe. Table 5 also shows ANOVA of MSPEs in the motor cortex. The variance ratios between the new method(1V) and the new method(8V), and those between the new method(1V) and the new method(27V) are greater than F-critical value. The variance ratios between the new method(8V) and the new method(27V) are less than F-critical value.

Table 4 and Table 5 show the results of the 200th(300th) regression formulas. At the nth regression formulas(n is neither 200 nor 300), we obtained similar results. We can conclude that the new method(8V or 27V) is more accurate than the new method(1V). As the regression ability of the conventional method(1V (cnv)) is almost the same as that of the new method(1V), we can also conclude that the new method(8V or 27V) is more accurate than the conventional method(1V (cnv)).

The accuracy of the new method (8V) is almost the same as that of the new method (27V). However, the averages of the MSPEs of the new method (8V) are a little less than the averages of the MSPEs of the new method (27V) (See Table 4). Therefore, in this paper, we use the new method(8V). We conjecture that the new method with $2^3$ or $3^3$ voxels may work well, and the new method with $n^3$ ($n$ is greater than 3) voxels may not work well, which is included in the future work.

### 3.2 The deactivations of interneurons

We performed two-tailed t-tests ($p \leqq 0.001$, FWHM=8mm) on the coefficients of the regression formulas in ascending order of their MSPEs until the number of significant voxels reaches a certain number, which is denoted by C hereinafter. We counted the number of positive significant voxels and the number of negative significant voxels. The deactivation ratio is calculated as follows:

$$\text{Deactivation ratio} = \frac{\text{the number of deactivated voxels}}{\text{the number of deactivated voxels} + \text{the number of activated voxels}}$$
$$= \frac{\text{the number of negative significant voxels}}{\text{the number of negative significant voxels} + \text{the number of positive significant voxels}}.$$

Table 6 shows the deactivation ratios (DRs) and MSPEs when C=200(300). The averages of the deactivation ratios in Table 6 are approximately 0.31-0.34. Notice that "200(300)" in Table 6 means the 200(300)th significant voxel, and "200(300)" in Table 4 means the 200(300)th regression formula. Generally, the 200(300)th significant voxels are included in the nth regression formulas (n>200(300)).

When C is less than 200 (for example, C=100), the statistical stability is small, that is, the deactivation ratios are unstable. When C is greater than 300 (for example, C=400), MSPEs are large, which means that the voxels included in the regression formulas are not related to the experimental task. Therefore, C=200(300) is appropriate. Generally, the appropriate C depends on the experimental task, that is, if the activation areas contain a lot of voxels, the appropriate C is large, and if the activation areas contain a small number of voxels, the appropriate C is small. It is another possible method to perform t-tests on the regression formulas whose MSPEs are less than a certain threshold (e.g. 0.15 or 0.20).

Table 6    The deactivation ratios with FWHM=8mm and p≦0.001

| No. | Temporal lobe | | | | Motor cortex | | | |
| --- | --- | --- | --- | --- | --- | --- | --- | --- |
| | C=200 | | C=300 | | C=200 | | C=300 | |
| | DR | MSPE | DR | MSPE | DR | MSPE | DR | MSPE |
| 1 | 0.343 | 0.105 | 0.357 | 0.120 | 0.365 | 0.127 | 0.357 | 0.146 |
| 2 | 0.280 | 0.119 | 0.283 | 0.139 | 0.345 | 0.118 | 0.327 | 0.145 |
| 3 | 0.290 | 0.110 | 0.307 | 0.128 | 0.350 | 0.148 | 0.360 | 0.168 |
| 4 | 0.310 | 0.107 | 0.280 | 0.132 | 0.390 | 0.150 | 0.363 | 0.171 |
| 5 | 0.340 | 0.135 | 0.302 | 0.167 | 0.340 | 0.153 | 0.329 | 0.175 |
| 6 | 0.330 | 0.0949 | 0.313 | 0.108 | 0.240 | 0.141 | 0.250 | 0.158 |
| 7 | 0.330 | 0.130 | 0.293 | 0.159 | 0.300 | 0.172 | 0.330 | 0.195 |
| 8 | 0.280 | 0.141 | 0.287 | 0.158 | 0.385 | 0.171 | 0.377 | 0.184 |
| 9 | 0.385 | 0.181 | 0.353 | 0.208 | 0.265 | 0.188 | 0.320 | 0.231 |
| 10 | 0.365 | 0.179 | 0.353 | 0.215 | 0.355 | 0.202 | 0.393 | 0.239 |
| Avg. | 0.325 | | 0.313 | | 0.334 | | 0.341 | |

"No." stands for the participant number. Notice that "200(300)" in Table 6 means the 200(300)th significant voxel, and "200(300)" in Table 4 means the 200(300)th regression formula. The averages of deactivation ratios are approximately 0.31-0.34.

We also analyzed the data with FWHM=6mm and p≦0.001. The average of deactivation ratios of the temporal lobe was 0.344, and that of the motor cortex was 0.344. Therefore the averages of deactivation ratios with FWHM=6mm and p≦0.001 are a little greater than those of Table 6(FWHM=8mm, p≦0.001). We also analyzed the data with FWHM=8mm and p≦ 0.0027, which corresponds to 3σ of the normal distribution. The average of deactivation ratios of the temporal lobe was 0.357, and that of the motor cortex was 0.366. Therefore the averages of deactivation ratios with FWHM=8mm and p≦0.0027 are a little greater than those of Table 6 (FWHM=8mm, p≦0.001). Hereinafter, let FWHM=8mm and p≦0.001, because they both are commonly used.

Let us investigate the deactivation ratios when C=300 in Table 6 in more detail. Table 7 shows the deactivation ratios (DRs) of the left hemisphere and the right hemisphere. The averages of deactivation ratios of the left hemisphere are almost the same as those of the right hemisphere. In the temporal lobe of No.5, left number=186 and right number=115, then 186+115=301, which is not equal to 300. The reason is that two voxels turned out to be significant by t-tests in the last regression formula.

**Table 7　The DRs of the left(right) hemisphere at the 300th significant voxel**

| No. | Temporal lobe | | | | Motor cortex | | | |
|---|---|---|---|---|---|---|---|---|
| | Left | | Right | | Left | | Right | |
| | Number | DR | Number | DR | Number | DR | Number | DR |
| 1 | 78 | 0.308 | 222 | 0.374 | 165 | 0.309 | 135 | 0.415 |
| 2 | 123 | 0.285 | 177 | 0.282 | 83 | 0.349 | 217 | 0.318 |
| 3 | 171 | 0.316 | 129 | 0.295 | 142 | 0.366 | 158 | 0.354 |
| 4 | 133 | 0.278 | 167 | 0.281 | 164 | 0.366 | 136 | 0.360 |
| 5 | 186 | 0.317 | 115 | 0.278 | 134 | 0.313 | 167 | 0.341 |
| 6 | 187 | 0.342 | 113 | 0.265 | 123 | 0.236 | 177 | 0.260 |
| 7 | 162 | 0.253 | 138 | 0.341 | 165 | 0.339 | 135 | 0.319 |
| 8 | 161 | 0.286 | 139 | 0.288 | 109 | 0.385 | 191 | 0.372 |
| 9 | 131 | 0.313 | 169 | 0.385 | 137 | 0.358 | 163 | 0.288 |
| 10 | 86 | 0.360 | 214 | 0.350 | 190 | 0.395 | 110 | 0.391 |
| Average | 141.8 | 0.306 | 158.3 | 0.314 | 141.2 | 0.342 | 158.9 | 0.342 |

"No." stands for the participant number. "Number" stands for the number of (positive and negative) significant voxels. The averages of deactivation ratios of the left hemisphere are almost the same as those of the right hemisphere.

We conducted the experiment of simple repetition five times with participant No.7. The deactivation ratios of the temporal lobe were 0.293, 0.363, 0.347, 0.310, and 0.347. The average was 0.332. The deactivation ratios of the motor cortex were 0.330, 0.387, 0.263, 0.317, and 0.307. The average was 0.321.

From Table 6 and Table 7, the averages of the deactivation ratios of the temporal lobe are approximately 30%-33%. The averages of the deactivation ratios of the motor cortex are approximately 33%-35%. It was reported that the ratio of the population of interneurons in the human temporal lobe is approximately 37.7%(7). In order to find out the relation between the deactivation ratios and the ratio of the population of interneurons, we have to consider the energy consumption of interneurons and the ratio of the deactivated interneurons in an activation area.

Let N stand for the ratio of the population of interneurons. Let A stand for the ratio of the average energy consumption of interneurons to the average energy consumption of excitatory neurons, which is called relative average energy consumption in this paper. Notice that the relative average energy consumption of excitatory neurons is 1.

$$A = \frac{\text{the average energy consumption of interneurons}}{\text{the average energy consumption of excitatory neurons}}.$$

Let B stand for the ratio of the population of deactivated interneurons to the population of interneurons in an activation area. For simplification, "in an activated area" is omitted hereinafter.

$$B = \frac{\text{the popluation of deactivated interneurons}}{\text{the population of interneurons}}.$$

Let F stand for the ratio of the energy consumption of deactivated interneurons to the energy consumption of the neurons. That is, F is described as follows:

$$F = \frac{\text{the energy consumption of deactivated interneurons}}{\text{the energy consumption of neurons}}.$$

As the neurons are divided into excitatory neurons and interneurons, F is as follows:

$$F = \frac{\text{the energy consumption of deactivated interneurons}}{\text{the energy consumption of interneurons} + \text{the energy consumption of excitatory neurons}}.$$

Let the numerator and the denominator of the above formula be divided by the energy consumption of excitatory neurons, then

$$F = \frac{\text{the relative energy consumption of deactivated interneurons}}{\text{the relative energy consumption of interneurons} + \text{the relative energy consumption of excitatory neurons}}.$$

The ratio of the population of deactivated interneurons is $B \times N$.

Let M stand for the total population of neurons in an activated area, then the population of deactivated interneurons is $B \times N \times M$.

The relative energy consumption of deactivated interneurons is $A \times B \times N \times M$.

The relative energy consumption of interneurons is $A \times N \times M$.

The relative energy consumption of excitatory neurons is $1 \times (1 - N) \times M$.

Therefore, F is as follows:

$$F = \frac{A \times B \times N \times M}{A \times N \times M + (1 - N) \times M} = \frac{A \times B \times N}{A \times N + 1 - N}.$$

Although the contributions of interneurons are very complicated at the micro level, the macro-level discussion is possible, because one voxel contains more than thousands of interneurons. Let us investigate A. The energy consumption of interneurons and those of excitatory neurons in humans are not well known. It was reported that the ratio of the population of interneurons in rats is approximately 15%(9,10,11). It was reported that the energy consumption ratio of interneurons in rats is 18%(12). Therefore, in rats, A is approximately 1.2. In humans, A may be higher than 1.2(13,14).

Let us investigate B. Interneurons are divided into Parvalbumin expressing (PV) interneurons, Somatostatin expressing (SOM) interneurons, Vasoactive intestinal peptide expressing (VIP) interneurons, and so on. When excitatory neurons in a certain area are activated, VIP interneurons are activated (excited) and inhibit other interneurons, and the interneurons inhibited by VIP interneurons are deactivated, which disinhibit excitatory neurons(15,16). Other inhibitions (for example, PV interneurons strongly inhibit one another) are also reported(17,18,19). Inhibition and disinhibition are controversial.

For further argument, some assumptions on A and B are needed. Let us assume that A of humans is the same as A of rats (A=1.2). Let us assume that VIP interneurons in an activation area are activated and the majority of the other interneurons in the activation area are deactivated. The ratio of the population of VIP interneurons to the population of interneurons is 9.7±1.0%(20). Therefore, let us assume that B is approximately $1 - 0.097 - \alpha (=0.903 - \alpha)$, where α stands for the ratio of activated interneurons other than VIP interneurons. Thus, A =1.2, B=0.903−α, and N=0.377. Let us assume that $0.1 \leqq \alpha \leqq 0.3$. Substitute the above values into $F = \frac{A \times B \times N}{A \times N + 1 - N}$, then $0.254 \leqq F \leqq 0.338$.

The averages of the deactivation ratios of the temporal lobe in Table 6 and 7 are approximately 30%-33%, and therefore the deactivations detected in the experiments are probably the deactivations of interneurons. Moreover, as A of humans is probably higher than A of rats(=1.2)(13,14), let us assume that A of humans is 1.5, for example. And let us assume that F is 0.313(the average of deactivation ratios of the temporal lobe when C=300 in Table 6), then

$$0.313 = \frac{1.5 \times B \times 0.377}{1.5 \times 0.377 + 1 - 0.377}.$$

holds, and B=0.658($\alpha$=0.245). Deactivated voxels and activated voxels are shown in Fig.2, which is explained in the next section.

### 3.3 The new method and the conventional method

We applied two conventional methods to the fMRI data of simple repetition. One is t-test ($p \leqq 0.001$). Another is t-test+ multiple comparison adjustment by random field theory, which is called "corrected" in Statistical Parametric Mapping (SPM)(5,21). Table 8 shows the deactivation ratios. The results of "1V" are similar to those of "0.05", because the regression formula of "1V" is obtained from that of "0.05" by exchanging the independent variables and the dependent variables. The deactivation ratios of "0.001" and "0.05" are approximately 0 with a few exceptions. The exceptions are No.1 temporal lobe, No.3 motor cortex, No.8 motor cortex and No.9 motor cortex.

Fig. 2 shows the activated voxels and the deactivated voxels of "0.05" and "8V" of the above four cases. In Fig.2, the averages from the brain surface to the 30th voxel are displayed using MRIcro(22). The activated voxels and the deactivated voxels of "0.05" are largely different from those of "8V". In "0.05", the activated voxels and the deactivated voxels basically are not adjacent. On the other hand, in "8V", the activated voxels and the deactivated voxels are adjacent. The (de)activated voxels in "8V" are less than those in "0.05", because the number of (de)activated voxels are limited to 300 in "8V". We can increase the number to 400, for example, but if we do so, the regression formulas with large MSPEs will be considered to be involved in (de)activations, which will degrade the quality of the results. We can set the limit to the number of voxels in "0.05". Even if we do so, the (de)activated voxels only will be less than those of Fig. 2, and the difference in the (de)activated voxels between "8V" and "0.05" will remain the same in essence. "Corrected" in the above argument is at the voxel-level, but at the cluster-level(5,21), the argument will be the same in essence.

Table 8    Deactivation ratios of four methods

| No. | Temporal lobe | | | | Motor cortex | | | |
|---|---|---|---|---|---|---|---|---|
| | 0.001 | 0.05 | 1V | 8V | 0.001 | 0.05 | 1V | 8V |
| 1 | 0.190 | 0.158 | 0.000 | 0.357 | 0.0393 | 0.0278 | 0.000 | 0.357 |
| 2 | 0.0526 | 0.0362 | 0.000 | 0.283 | 0.000 | 0.000 | 0.000 | 0.327 |
| 3 | 0.0703 | 0.0693 | 0.000 | 0.307 | 0.386 | 0.307 | 0.020 | 0.360 |
| 4 | 0.00270 | 0.000 | 0.000 | 0.280 | 0.000 | 0.000 | 0.000 | 0.363 |
| 5 | 0.0181 | 0.000 | 0.000 | 0.302 | 0.0320 | 0.000 | 0.000 | 0.329 |
| 6 | 0.0183 | 0.00802 | 0.000 | 0.313 | 0.00896 | 0.000 | 0.000 | 0.250 |
| 7 | 0.00866 | 0.000621 | 0.000 | 0.293 | 0.0158 | 0.000 | 0.000 | 0.330 |
| 8 | 0.0285 | 0.0108 | 0.000 | 0.287 | 0.144 | 0.108 | 0.0233 | 0.377 |
| 9 | 0.0266 | 0.00483 | 0.000 | 0.353 | 0.260 | 0.148 | 0.000 | 0.320 |
| 10 | 0.0174 | 0.00184 | 0.000 | 0.353 | 0.000 | 0.000 | 0.000 | 0.393 |

"No." stands for the participant number. "0.001" stands for t-test ($p \leqq 0.001$)(two-tailed). "0.05" stands for "corrected" ($p \leqq 0.05$)(one-tailed). "1V" stands for the new method(1 voxel) when C=300. "8V" stands for the new method(8 voxels) when C=300. The results of "1V" are similar to those of "0.05", because the regression formula of "1V" is obtained from that of "0.05" by exchanging the independent variable and the dependent variable. The deactivation ratios of "0.001" and "0.05" are approximately 0 with a few exceptions. The exceptions are No.1 temporal lobe, No.3 motor cortex, No.8 motor cortex and No.9 motor cortex.

The results of the new method(8V) are more reliable than those of the conventional method, because the new method(8V) is more accurate than the conventional method. Moreover, the deactivation ratios of the new method(8V) are stable, while the deactivation ratios of the conventional method are unstable, and the deactivations detected by the new method(8V) can be reasonably interpreted as the deactivations of interneurons, while the deactivations detected by the conventional method can be hardly interpreted. Many results on the deactivations (NBRs)(23,24,25) were obtained by the conventional method. The results by the new method(8V) will be probably different from those by the conventional method.

| No. | Area | Method | Left Activated voxels | Left Deactivated voxels | Right Activated voxels | Right Deactivated voxels |
|---|---|---|---|---|---|---|
| 1 | Temporal Lobe | 0.05 | | | | |
| 1 | Temporal Lobe | 8V | | | | |
| 3 | Motor cortex | 0.05 | | | | |
| 3 | Motor cortex | 8V | | | | |
| 8 | Motor cortex | 0.05 | | | | |
| 8 | Motor cortex | 8V | | | | |
| 9 | Motor cortex | 0.05 | | | | |
| 9 | Motor cortex | 8V | | | | |

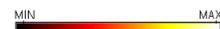

**Figure 2   The activated voxels and the deactivated voxels of the conventional method**

**and the new method**

"No." stands for the participant number. "0.05" stands for "corrected" (p≦ 0.05). "8V" stands for the new method(8 voxels) when C=300. The (de)activated voxels of "0.05" are largely different from those of "8V". For example, in the left deactivated voxels of participant 1 and participant 9, and in the right deactivated voxels of participant 3 and participant 8, the results of the two methods are largely different.

## 4. Discussion

We have presented a new fMRI data analysis method, which is more accurate than the conventional method. The deactivation ratios of the temporal lobe obtained by the new method are approximately 30%-33%, which probably mean the deactivations of interneurons by considering the ratio of the population of interneurons, the energy consumption of interneurons, and the inhibition and the disinhibition of interneurons. The deactivation ratios of the motor cortex are almost the same as those of the temporal lobe. The energy consumption of interneurons in humans are not well known. The inhibition and the disinhibition of interneurons are controversial. Therefore, in this paper, we have carried the argument on deactivation ratios under the obscure information. However, in the future, the argument will be more precise.

Those who had the health problems such as high blood pressure, low blood pressure, lack of sleep, fatigue, and hunger, and those who take medicines (especially, the medicines that affect the cardiovascular system) were excluded from the experiments. However, as a reference, we also conducted the experiments with the above people. The deactivation ratios of the above people were higher (approximately 40%-60%) than Table 6. The reason may be local blood stealing, wherein blood is diverted to neurally active regions without a concomitant change of neural activity in the negative BOLD regions(26), because those who have the health problems may be unable to supply the sufficient blood to the activation areas, and the vascular systems of those who take medicines may not work well due to the medicines. The deactivation ratios of those who have the health problems or take medicines may equal the deactivations of interneurons plus local blood stealing. Participant No.9 had a little high blood pressure, and participant No.10 had a little low blood pressure. Therefore, the deactivation ratios of No.9 temporal lobe, No.10 temporal lobe and No.10 motor cortex are a little large among the deactivation ratios (See Table 6). Moreover, their MSPEs are a little greater than those of the others (See Table 6), perhaps because blood stealing does not occur synchronously with tasks/rests, which increases their MSPEs.

We conducted the experiments of simple repetition which activates the temporal lobe. We also conducted other experimental tasks which activate the temporal lobe. As far as we experienced, when the activation areas of an experimental task were large, the deactivation ratios tended to get larger. The reason may be that experimental tasks whose activation areas are large need a lot of blood, which easily causes local blood stealing. Therefore experimental tasks whose activation areas are large may be inappropriate for detecting the deactivations of interneurons.

The (de)activated voxels detected by the conventional method, which are largely different from those by the new method(8V), may be incorrect, because the regression formulas of the conventional method are less accurate than those of the new method(8V). Moreover, in the conventional method, right-tailed t-tests are usually performed, which is based on the idea that only the excitations of excitatory neurons are the activations. That is, right-tailed t-tests neglect the deactivations of inhibitory interneurons. However, a large portion of the deactivations of inhibitory interneurons is also considered to be activations. Therefore, the right-tailed t-test does not detect the whole activation. A lot of fMRI studies thus far by the conventional method should be re-examined by the new method, and many results obtained thus far will be modified.

Interneurons in the human cortex can hardly be investigated due to several problems (e.g. ethical problems). As the new method can detect the deactivations of interneurons, the new method will be a powerful tool for the study of the interneurons in the human cortex. The loss or dysfunction of interneurons is related to several diseases such as epilepsy, Alzheimer's disease, schizophrenia and so on(27,28,29). As the new method can detect the deactivations of interneurons, the new method will be a powerful tool for the image diagnosis of the above diseases.

## 5. Methods
### 5.1 Participants

Ten participants in the experiments were drawn from right-handed healthy volunteers with normal blood pressures. Those who had the health problems such as lack of sleep, fatigue, and hunger and those who took medicines (especially, the medicines that affect the cardiovascular system) were excluded. Eight participants were male in their twenties. One participant was male in his sixties. One participant was female in her twenties. The study was approved by the local ethics committee, and complied with the relevant ethical regulations. Written informed consents were obtained from all participants.

**5.2 fMRI measurements**

fMRI measurements were performed with a 1.5T MRI system. The measurement parameters were as follows: slice thickness: 4mm, slice gap: 1mm, slice number: 24, matrix size: 64×64, echo time: 45msec, repetition time: 6000msec, flip angle: 90°, field of view: 240mm×240mm. The block design consisted of six task blocks and six rest blocks. Each block consisted of eight volumes (96 volumes in total).

**5.3 Data analysis**

In Section 2 and Section 3, we explained the main points. We explain what we did not mention in Section 2 and Section3.

1. Preprocessing included realignment, normalization and smoothing (full width at half maximum (FWHM) = 8mm or 6mm) using SPM8(21) . Tasks/rests were convolved with HRF.

2. When cubes with a side of 2 voxels are generated in the temporal lobe, some cubes consist of the voxels inside the temporal lobe and the voxels outside the temporal lobe. We performed the regression analysis of the cubes consisting of the voxels inside the temporal lobe and the voxels outside the temporal lobe. We performed t-tests on the coefficients inside the temporal lobe, but we did not perform t-tests on the coefficients outside the temporal lobe. We did the same treatment in the motor cortex.

3. In a cross validation, regression analyses are performed with 95(=96－1) volumes 96 times, and 96 different regression formulas are generated. Which regression formula should be used for t-test? There is no special reason to select a certain regression formula among them for t-test. Therefore, regression analysis with 96 volumes was also performed to generate the regression formula for t-test.

4. T-tests were performed until the number of (de)activated voxels reaches a certain predetermined number (for example, 300). Sometimes, the number of (de)activated voxels did not reach the predetermined number. For example, when a participant did not perform an experimental task appropriately, the regression formula obtained was inaccurate, and the number of (de)activated voxels did not reach the predetermined number.

**Acknowledgements**

We thank Hiroyuki Hioki for teaching interneurons. We thank the participants in the experiments.